# The Growing US-Mexico Natural Gas Trade and Its Regional Economic Impacts in Mexico

*Haoying Wang*
(haoying.wang@nmt.edu)

Department of Business and Technology Management, NM Tech

*Rafael Garduño Rivera*
(rgardunor@up.edu.mx)

Universidad Panamericana Campus Aguascalientes

**Abstract:**

With the recent administration change in Mexico, the fluctuations in national energy policy have generated widespread concerns among investors and the public. The debate centers around Mexico's energy dependence on the US and how Mexico's energy development should move forward. The goal of this study is two-fold. We first review the history and background of the recent energy reforms in Mexico. The focus of the study is on quantifying the state-level regional economic impact of the growing US-Mexico natural gas trade in Mexico. We examine both the quantity effect (impact of import volume) and the price effect (impact of natural gas price changes). Our empirical analysis adopts a fixed-effects regression model and the instrumental variables (IV) estimation approach to address spatial heterogeneities and the potential endogeneity associated with natural gas import. The quantity effect analysis suggests a statistically significant positive employment impact of imports in non-mining sectors. The impact in the mining sector, however, is insignificant. The state-level average (non-mining) employment impact is 127 jobs per million MCFs of natural gas imported from the US. The price effect analysis suggests a statistically significant positive employment impact of price increases in the mining sector. A one-percentage increase in natural gas price (1.82 Pesos/GJ, in 2015 Peso) leads to an average state-level mining employment increase of 140 (or 2.38%). We also explored the implications of our findings for Mexico's energy policy, trade policy, and energy security.

# 1. Introduction

With a total population of around 450 million together, Mexico and the US consume lots of energy annually. In 2018 alone, the US energy consumption was about 101 quadrillion BTUs, and Mexico consumed about 8 quadrillion BTUs according to the US Energy Information Administration (US EIA). A substantial portion of this demand was met by oil and natural gas. On the supply side, the energy landscape is very different. While the US has become a net energy exporter in recent years following a surge in production since the mid-2000s (Figure 1, left), Mexico has been an energy importer for a long time, with the US being its main trade partner (Figure 1, right). Mexico's energy dependence on natural gas imports can be revealed through the natural gas price dynamics. As Figure 2 (left panel) shows, Mexico's domestic natural gas retail price closely follows the US export price for Mexico. Moreover, Mexico's import need for natural gas has been rising as domestic production stagnates and demand increases, particularly in the electricity sector (Navarro-Pineda et al., 2017). Mexico has been importing most of its natural gas supply from the US (Figure 2, right). The consequences of the growing energy trade in Mexico have been understudied. A few recent studies have explored this issue from the perspective of Mexico's energy independence and security (e.g., Baker, 2016; Vietor and Sheldahl-Thomason, 2017). To the best of our knowledge, there has been no study systematically examining the regional economic impacts of the growing US-Mexico natural gas trade in Mexico. With the recent administration change and the possible slowdown of the energy reform process, the role of the energy sector in Mexico's economy has been in the debate (Graham, 2020). This study seeks to understand the regional economic impacts of the growing US-Mexico natural gas trade. Our focus is on the past two decades that mostly correspond to the recent three administrations (1998-2019), during which Mexico enjoyed stable economic growth and its GDP increased by more than 50%. Several major energy policy reforms also emerged during the period.

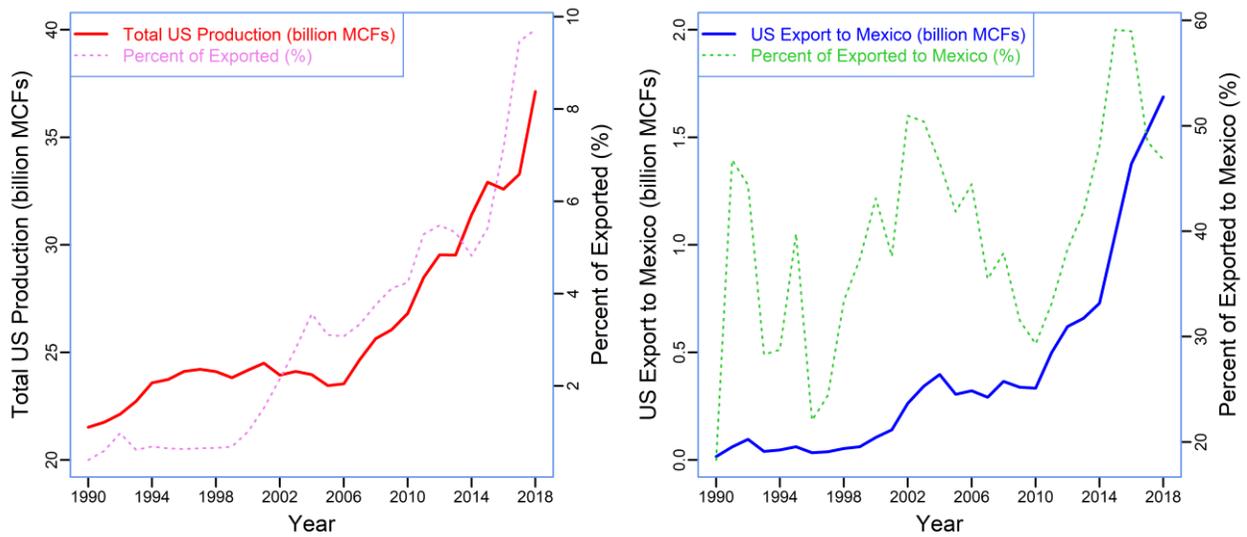

Data source: US EIA.

**Figure 1.** US natural gas production and export (left: to all countries; right: to Mexico only)

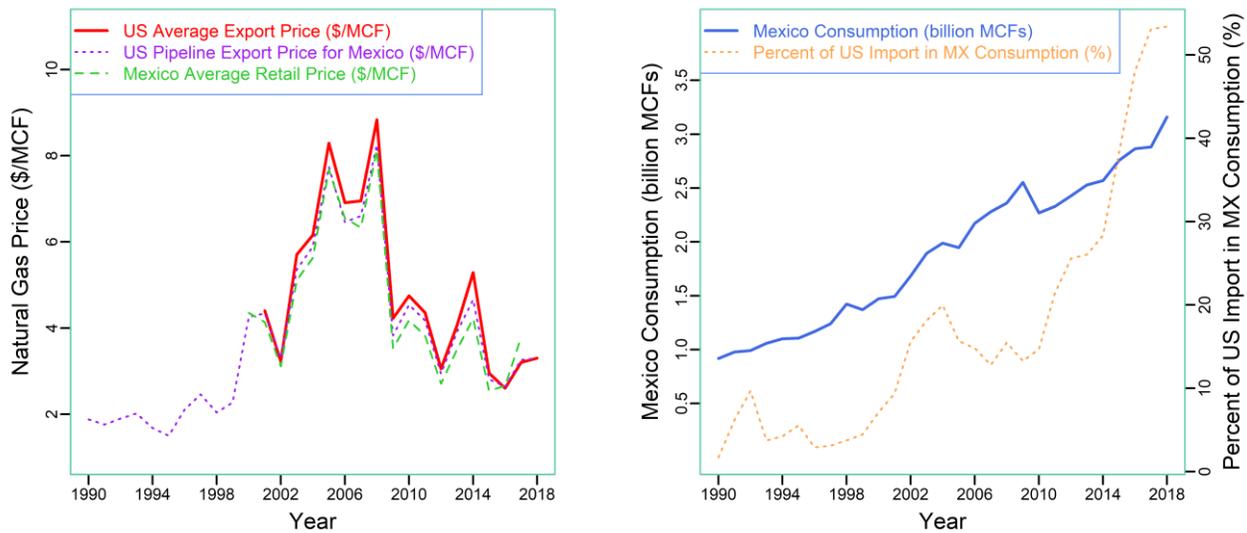

Data source: US EIA, BP Statistical Review of World Energy 2019, Mexican Minister of Energy (SENER). Note: The systematic difference between US export prices and Mexico's average retail prices is likely a result of currency conversion. The Mexico average retail price is expressed in Peso in the raw data. It is converted into US Dollars using the historical MXN-USD exchange rates.

**Figure 2.** US natural gas export prices and Mexico natural gas consumption and retail price

Being different from many other trade partnerships, Mexico and the US share an almost 3200km border, which has facilitated a long history of cross-border trade of commodities. The US-Mexico natural gas trade has grown rapidly in the past two decades following the expansion of cross-border natural gas pipelines. Much of the pipeline capacity started operating in the last decade. Figure 3 shows the major natural gas pipeline cross points along the US-Mexico border. Most of the new capacity expansion follows the recent shale development in the Permian Basin (Western Texas and Southeastern New Mexico) and the Eagle Ford Shale region (Southern Texas). The low natural gas price after 2009 was one of the main drivers of the expanding natural gas trade (Figure 2, left). It is worth noting that the low natural gas price is mostly an endogenous outcome of several shale formation discoveries in the US, particularly the Marcellus Shale in Pennsylvania, Ohio, and West Virginia.

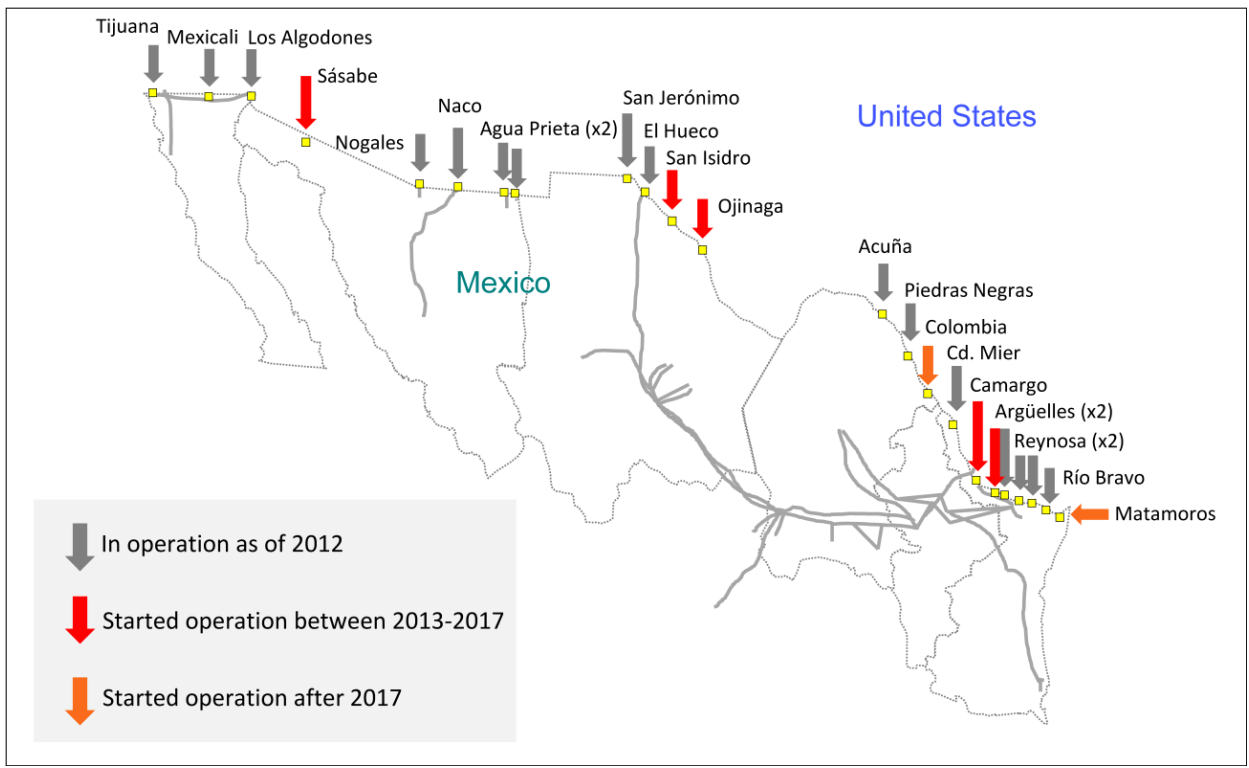

Data source: Mexican Ministry of Energy (Secretaría de Energía, or SENER)

**Figure 3.** Locations of natural gas pipelines in the US-Mexico transboundary area

Mexico's new energy reform (2013 – 2014) is another driver of the growing natural gas trade. The reform changed the governing system of Mexico's natural gas sector by bringing private and foreign investment into the energy market to improve infrastructure and energy security (Ibarzábal, 2017). For the first time after World War II, Mexico's energy market was open to private competition. Market price becomes an effective signal for natural gas supply and demand. Driven by both the supply side shock (i.e., the shale boom in the US) and the energy market structure change, Mexico's domestic natural gas pipeline network has expanded substantially in recent years. Figure A1 in the Appendix illustrates Mexico's domestic natural gas pipeline network expansion after the new energy reform. Figure 4 shows the growth of the domestic pipeline network (right panel) and pipeline capacity across the US-Mexico border (left panel). The growing natural gas trade and the expanding pipeline network have raised considerable debates over their social, economic, and environmental impacts in Mexico (Navarro-Pineda et al., 2017; Ibarzábal, 2017; Russo, 2017; Fine and Loris, 2019). For instance, concerns over policy fluctuation between different administrations and the impact on the natural gas sector have been raised (Fine and Loris, 2019). Ibarzábal (2017) argued that Mexico's natural gas transmission pipeline system is difficult to govern because of its high complexity, and the recent energy reform may have worsened the situation. This study focuses on understanding the regional economic impacts of the growing US-Mexico natural gas trade in Mexico. Related to the regional economic impacts, we also explore the relevant policy implications for Mexico.

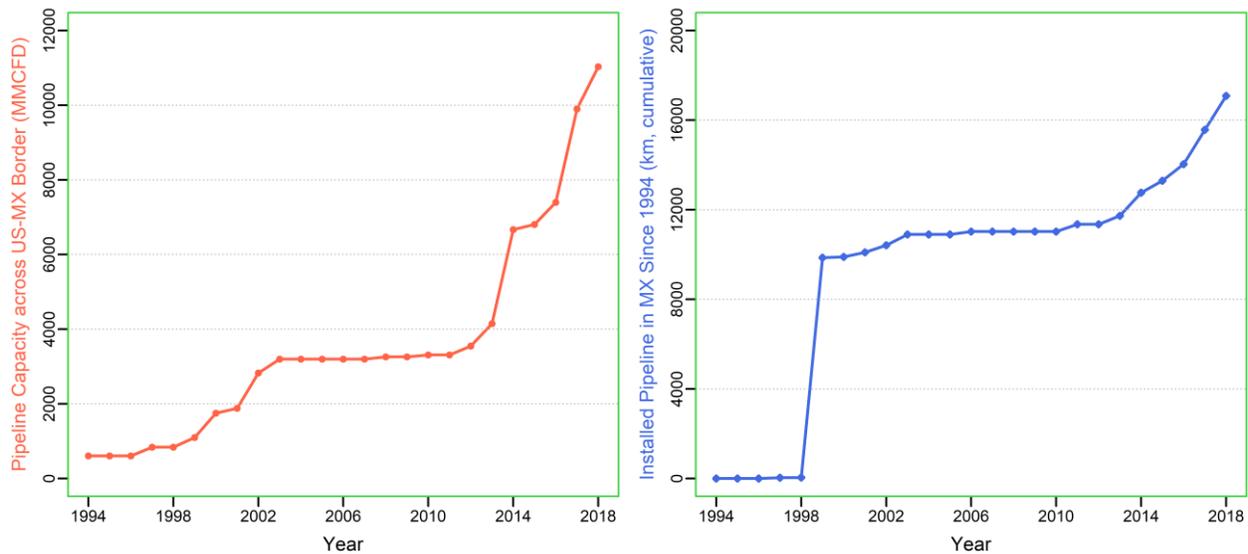

Data source: US EIA and Mexican Ministry of Energy (Secretaría de Energía, or SENER).

**Figure 4.** Natural gas pipeline capacity crossing the US-Mexico border and the total length in Mexico

## 2. History and Policy Background

Mexico discovered its first natural gas deposit in Misión, Northern Mexico in the 1940s. Due to difficulties that PEMEX (Mexican Petroleum, the Mexican state-owned petroleum company) confronted the following years, however, it was until a decade later that PEMEX started exploiting the natural gas. The extraction increased during the 1950s and 1960s when it grew from 0.256 to 1.325 billion cubic feet per day (Bcf/D) (Márquez, 1988). By 2009 Mexico reached its all-time high production, but the production level has decreased ever since and reduced to half of the 2009 peak level in 2018.[1] Despite the decline, British Petroleum ranked Mexico as the 13th world's largest gas producer and the 11th largest oil producer in 2018. Nevertheless, the US natural gas supply has grown so rapidly in recent years that Mexico has chosen to decrease its extraction and import from the US instead.

Meanwhile, Mexico's energy sector has gone through important reforms. Multiple structural changes in different sectors reshaped Mexico's economy in the last three decades (Padilla-Pérez and Villarreal, 2017). The 1990s saw a surge of privatization across several sectors such as telecommunication and the steel industry. Other sectors were deregulated and opened to licensing, including seaport services and storage & transportation of natural gas (OECD, 2004). It was not until the 2000s that the energy sector was transformed. Two important sets of reforms occurred in the 2000s. First, the Calderon administration (2006-2012) implemented five reforms (fiscal, pension system, energy, competition, and labor reforms). Then, the Peña-Nieto administration (2012-2018) implemented 16 reforms, of which 11 corresponded to the "Pact for Mexico" (Gutiérrez, 2014; OECD, 2015; Zorrilla, 2017). Mexican Energetic Reform approved in 2013 was dedicated to energy issues. It allows the acquisition of electricity under competitive prices in the wholesale electricity market.

---

[1] See https://www.ceicdata.com/en/indicator/mexico/natural-gas-production-opec-marketed-production, accessed on April 2, 2020.

The most important reforms during the Calderon administration were energy-related. Its energy reform focused on legislative changes, including the bidding process allowing oil fields open to private enterprises, the Comprehensive Exploration and Production Contracts (CIEP) initiative, and the signing of an agreement with the US to exploit hydrocarbons in cross-border deposits. However, these reforms did not improve the sector's production and technology capacity. They tend to be ad hoc bureaucratic and administrative reforms. The reforms led to more high-level bureaucracy and red tape. They also increased the public expenditure and PEMEX´s debt level. Among several efforts, the creation of the Energy Regulatory Commission stood out (Gutierrez, 2014). To some extent, these reforms tried to replicate the success achieved in the US oil & gas sector. It is worth pointing out that, even if Mexico's natural gas deposits are similar to those in the US, their geologic conditions are more complicated making them difficult to develop. Most of the natural gas deposits in Mexico are around 5km deep (Cordano and Zellou, 2020).

The "Pact for Mexico" drove most of the structural reforms during the Peña-Nieto administration. The Pact was a political agreement between the Institutional Revolutionary Party (PRI, the President's party), the National Action Party (PAN), and the Party of the Democratic Revolution (PRD) that seeks to address security problems, poverty, inequality, corruption, and to promote economic growth (Zorrilla, 2017).

These reforms occurred in the international context of an oil price decrease and a more restrictive US monetary policy. Still, the prospect was that the reforms would have a positive impact on Mexico's economic growth (OECD, 2015 and 2017). Apart from the "Pact for Mexico" reform, the Mexican government planned five more reforms. From 2012 to 2015, Mexico had the most economic reforms among all OECD countries (OECD, 2015). Nevertheless, natural gas production in Mexico has not been able to keep up with the increasing demand since 2011: production fell on average 0.2% yearly while demand grew around 4.3% yearly (Cordano and Zellou, 2020). Meantime, energy demand saw the opportunity to switch from other energy sources (i.e., coal and oil) to the cheaper and cleaner alternative - natural gas. In the end, these reforms did not boost the supply as expected. And Mexico became more dependent on US natural gas export.

A radical energy market reform started in 2013 and expected an investment between $175 and $200 billion US dollars to create around 70 new energy firms (World Bank, 2019, pp. 61-72). The reform pursued mainly two things related to hydrocarbons. First, it reduced government control that has predominated since the Cardenas administration (1934-1940). The entire energy sector was nationalized back then. Second, it incentivized industry modernization through capital and technology investment. The Mexican government intended to attract direct investment through these changes. For instance, the reform deregulated the chemical industry in 2014, which allowed the industry to benefit from the US shale boom.

In 2015, Mexico started its bidding process for the exploration and extraction of hydrocarbons. There were three bidding rounds by the end of 2015. However, even with successful biddings, there was no commitment to a substantial investment. Thus, Mexico has not been able to increase its natural gas production while having a significant deficit with the US for over 50% of its natural gas consumption (Figure 2, right). As a solution to the investment problem, Mexico initiated its strategy to exploit the shale gas recourses. However, the Mexican Energy Ministry (SENER) has postponed the exploration and exploitation "until new notice" to have time to review the information from the bidding firms (Cordano and Zellou, 2020). It is still necessary to see if the reforms are successful or not in practice. Moreover, the current López-Obrador administration has decided to review all the bidding rounds from the previous administration to see if they were assigned lawfully, which adds more uncertainty. For instance, the current administration might decide that the bidding rounds should be redone after the review. These uncertainties send a negative signal to potential investors as they likely learn from the history of how private corporate assets were expropriated during the Cardenas administration years.

Overall, the energy reforms have set the market foundation for the future growth of Mexico's energy sector. There are still policy objectives to achieve, such as fostering competitiveness, improving market efficiency and technology innovation, and attracting much-needed investment (World Bank, 2019, pp 61-72). Mexico's oil and gas industry has thus far failed to reach the expected outcome of the reforms. Therefore, it is reasonable to expect that Mexico will continue relying on US natural gas export in the foreseeable near-to-medium term. Next, we quantify the regional economic impacts of such an energy dependence.

## 3. The Regional Economic Impact

*3.1 A Brief Literature Review*

Due to the increase of drilling in the US and the fall of natural gas prices, Mexico will likely continue to depend on US natural gas instead of extracting its own. Mexico has discovered shale gas deposits in the past years, including one of the world's largest reserves along the US-Mexico border in Tamaulipas (Comisión Nacional de Hidrocarburos, 2018). It is estimated that the new natural gas reserve will produce 545 billion MCFs (Brasier and Thompson, 2017). One of the main reasons for Mexico's natural gas dependence on imports is the high drilling cost, as mentioned previously. Thus far, Mexico and the US have focused on building more pipelines and facilitating the energy trade rather than promoting investments in developing the natural gas deposits in Mexico. The debate between relying on cheap imported natural gas and establishing energy independence is ongoing.

A few recent studies have emphasized the importance of energy independence for Mexico (e.g., Cordano and Zellou, 2020; Laguna-Martinez et al., 2020). They all agree that Mexico has not taken real action to pursue energy independence. For instance, Laguna-Martinez et al. (2020) concluded that shale gas development could establish energy independence for Mexico. They also stressed that the lack of fresh water in some areas is a critical factor limiting natural gas exploitation. At the same time, the current López-Obrador administration focuses on energy sovereignty while promising environmental protection. Its energy policies are not necessarily going to create a new energy system with lower prices for Mexican consumers and better environmental performance (e.g., reduced greenhouse gas emission) (Gross, 2019).

Another factor to consider is Mexico's socio-economic insecurity. In the last four decades, different drug cartels have controlled the regions along the US-MX border (Haahr, 2015). If the Mexican government commences exploring a natural gas reserve, the drug cartels will ask for a quota or the control of the drilling sites.[2] The situation is the worst in the Burgos region of Tamaulipas, where the largest natural gas reserve was discovered. Tamaulipas is also one of the states with more drug homicides in the last decade.[3] The state is seriously affected by the Gulf cartel who will not allow any profitable business to operate without being part of it (Haahr, 2015). Therefore, it is likely a contributing factor for why the Mexican government has opted to continue importing natural gas from the US instead of developing its reserve.

Overall, the consensus in the literature is that depending on US natural gas supply makes Mexico vulnerable to natural gas market volatility and US energy policy. As suggested by González (2016), while Mexico continues taking advantage of cheap US natural gas, it should also develop infrastructure, technology, and business environment to prepare for a boost in the natural gas sector. All these can be integrated into an environmental, legal, and economic framework to transform the natural gas sector by leveraging the natural gas deposits discovered in Mexico. It will also provide new opportunities for employment in the natural gas sector and beyond. While this seems to be a reasonable long-term economic development strategy, a policy-relevant question is: what is the near-to-medium-term economic impact of importing natural gas from the US?

---

[2] For instance, see https://www.reuters.com/article/us-mexico-drugs-energy/mexican-drug-gangsters-menace-natural-gas-drillers-idUSTRE71E4GY20110215, accessed April 1, 2020.
[3] See https://vanguardia.com.mx/articulo/cartel-del-golfo-y-zetas-ahuyentan-fracking-en-tamaulipas, accessed April 1, 2020.

The classical trade theory suggests that comparative advantages can bring mutual benefits to trade partners. In the case of natural gas, the US has a comparative advantage. Hence, the theory predicts that the regional economies in Mexico can benefit from importing US natural gas, aside from the aforementioned energy independence issue. When it comes to policymaking, energy dependence can be a strategic part of long-term economic development planning (Bluszcz, 2017). The literature on the economic impact of Mexico's growing natural gas import is limited. Most studies focus on issues related to the energy trade deficit and the energy security debate. Dávila Flores (2013) is the only relevant economic impact study that we can find. It shows that the average wage in the natural gas sector is significantly higher in Northeastern Mexico. The region is also where many cross-border natural gas pipelines pass through (Figure 3). In the broader literature, Coronado and Zellou (2020) emphasized the importance of shale gas extraction for regional economic development in Latin America. However, the lack of investment in shale development in Mexico has not propelled job growth and economic prosperity as expected. One of the reactivation plans by the current López-Obrador administration is to restart the projects that were left behind years ago, for instance, the construction of new natural gas pipelines in Salina Cruz and Coatzacoalcos (Grayson, 1981). The new government expenditure can increase employment in natural gas and related sectors. As of this writing, however, we have not seen any revival of these projects.

In the following subsections, we focus on the employment impact of Mexico's growing natural gas import from the US. We adopt a regression analysis framework to estimate the impacts of natural gas import and price changes on employment in both the mining sector and all non-mining sectors. We take an instrumental variable (IV) estimation approach to address the potential endogeneity issue concerning natural gas import.

*3.2 Descriptive Trends in Employment*

Being different from the US, where there has been a significant job creation associated with shale gas development, Mexico has seen a decline in employment in the natural gas sector. The domestic supply has reduced in recent years. Table 1 shows the total (nationwide) employment and employment in the oil and gas extraction subsector (Code 211110) for the census years 2004, 2009, and 2014 in Mexico. There are three limitations associated with the information provided by the National Institute of Statistics and Geography (INEGI) of Mexico. First, the economic Census happens every five years. It is difficult to assess how employment changes annually. Second, the information of the latest Census (2019) has not been released for the oil and gas extraction subsector, making it impossible to see how its employment has changed after 2014, especially considering the impact of the new energy reform in 2013. Third, the statistics aggregate the oil and gas extraction subsectors. Therefore, it is difficult to observe how the employment of the natural gas industry alone has changed. Aside from these limitations, Table 1 suggests the number of jobs in the oil and gas subsector has declined as a percentage of the total employment nationwide. Although there was an increase from 46,652 jobs in 2004 to 53,581 jobs in 2014, the percentage reduced by almost 0.04% (or a 13.6% decrease). Overall, by steering the growing demand to US natural gas instead of developing its natural gas industry, one of the costs to Mexico is an overall employment decline in the oil and gas subsector. Some of the secondary sectors have been growing, such as pipeline construction in the US-Mexico border region, which increases Mexico's dependence on US natural gas (USDOE, 2020).

**Table 1.** Mexico total employment and employment in the oil and gas extraction sub-sector

| Year | Economic activity (sector) | Total jobs | % of oil & gas extraction jobs |
|---|---|---|---|
| 2014 | National total | 21,576,358 | 100% |
| | Oil and gas extraction (211110) | 53,581 | 0.248% |
| 2009 | National total | 20,116,834 | 100% |
| | Oil and gas extraction (211110) | 50,273 | 0.250% |
| 2004 | National total | 16,239,536 | 100% |
| | Oil and gas extraction (211110) | 46,652 | 0.287% |

Data source: INEGI (National Institute of Statistics and Geography), Mexico.

*3.3 A Regression Model of Regional Employment*

We start with a regression model with state fixed effects to estimate the impact of natural gas import on state-level employment (in the mining sector and non-mining sectors). The dependent variable is the annual employment count in a given sector (*EMP*). Independent variables include annual natural gas import from the US (*NGI*), annual population estimate at the state level (*POP*), and the Euclidian distance to the US-Mexico border (*DIST*) from the given state. It is worth noting that annual natural gas import does not have state-level variations. Hence, we cannot include year fixed effects in the model to absorb any temporal trends. Instead, we use state-level population to control the temporal trends as population and employment are highly correlated. In addition, we cannot include a stand-alone distance to the border variable because the model already controls for state-level fixed effects. Specifically, we estimate the following regression model as the baseline ($i$ and $t$ are the indices for state and year, respectively):

$$EMP_{it} = \beta_1 POP_{it} + \beta_2 NGI_t + \beta_3 (NGI_t \times DIST_i) + \mu_i + \varepsilon_{it} \qquad (1)$$

where $\mu_i$ represents state-level fixed effects to implicitly control any spatial heterogeneities unique to each state. $\varepsilon_{it}$ is the error term capturing any random shocks to employment. $\beta_1$ is the parameter associated with population. $\beta_2$ and $\beta_3$ are the parameters of interest. Their estimates allow us to derive a state-specific average employment impact of natural gas import. An empirical concern for the model in equation (1) is the potential endogeneity. In this study, a common factor that simultaneously drives both employment and natural gas import can cause an endogeneity issue. The estimate for $\beta_2$ will then be biased. To address the issue, we use a two-stage least squares (2SLS) IV regression. The methodological details will be discussed in the next sub-section.

The proposed regression model in equation (1) explores the quantity effect of the natural gas trade. An alternative channel of regional economic impact is through the price effect. If the natural gas demand is inelastic (to price changes) in Mexico, the quantity effect dominates. However, if the natural gas demand is elastic (e.g., substitutable with other energy sources), then the price effect may become important. In practice, we expect the price effect to be important in most of the non-mining sectors, where price is the main channel of impact on employment. In the mining subsectors, especially those related to natural gas, and subsectors concerning pipeline construction and operation, the quantity effects are expected to be more important. The following analysis will examine the mining sector and non-mining sectors separately. Figure 2 (left panel) shows that Mexico's domestic average natural gas retail price closely follows the US natural gas export price. The almost perfect correlation between retail price and US export price implies that Mexico's domestic natural gas price is driven by exogenous variations. The exogeneity allows us to examine the effect of price change on employment with a standard fixed effects panel data model similar to equation (1). Let $NGP_{it}$ denotes state-level natural gas price, we have the following regression model:

$$EMP_{it} = \beta_1 POP_{it} + \beta_2 NGP_{it} + \beta_3 (NGP_{it} \times DIST_i) + \mu_i + \varepsilon_{it} \qquad (2)$$

Now the key independent variable $NGP_{it}$ has both spatial and temporal variations, which is better for model identification. Although the endogeneity of natural gas price is not a concern here, we still need to control the state-level dependency on natural gas. Ideally, we can use the historical natural gas consumption level around 2000 (the beginning of our study period). The historical state-level natural gas consumption data in Mexico, however, is unavailable. Instead, we again use the distance to the US-Mexico border (*DIST*) as a proxy for the dependency on natural gas. Over the past two decades, Mexico has become more dependent on importing natural gas from the US through pipelines (Figure 2). Therefore, being closer to the border gives a cost advantage. It is also evident from the state-level changes in natural gas retail price. Figure 5 shows that most of the states in northern Mexico and central Mexico (except for Sonora and Durango) have seen a decline in natural gas prices (in real term) in the past two decades compared to southern Mexico. It suggests that being geographically closer to the US brings greater industrial integration and economic impacts (through the price effect). In addition, the natural gas pipeline network expands from north to south, as indicated in Figure A1. It implies that being closer to the border also has an access advantage and likely a higher dependency on natural gas. In the following two subsections, we explore the quantity effect ($\beta_2$ and $\beta_3$ in equation (1)) and the price effect ($\beta_2$ and $\beta_3$ in equation (2)) of the natural gas trade separately.

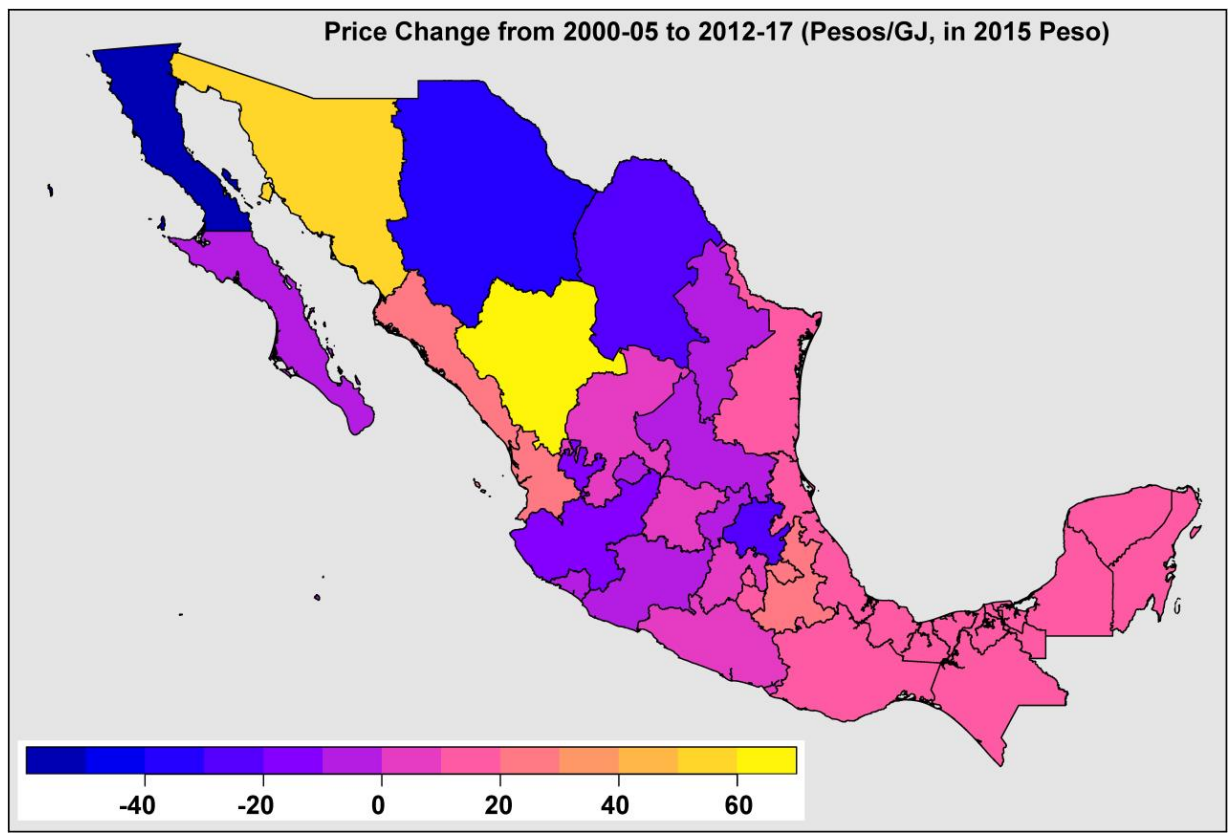

Data source: computed by the authors with information from the Energy Regulatory Commission (CRE, SENER) (SENER, 2021)

**Figure 5.** State-level changes of natural gas retail price from 2000-05 to 2012-17

*3.4 Quantity Effect - data and empirical results*

We assemble data for the regression analysis from different sources. The natural gas import data (both volume and price) come from the US EIA. We use the price information to derive instrumental variables for the IV regression. The state-level population and employment data come from the National Institute of Statistics and Geography (INEGI) of Mexico, specifically, the National Survey of Occupation and Employment (ENOE). The (Euclidean) distance to the border is measured from the geographic center of each state to the US-Mexico border using ArcGIS. It is worth noting that, for states south of Mexico City, the distance to the US-Mexico border is computed as the distance to Mexico City plus the distance from Mexico City to the US-Mexico border. These states include Guerrero, Morelos, Puebla, Tlaxcala, Veracruz, Oaxaca, Chiapas, Tabasco, Campeche, Quintana Roo, Yucatán. Table 2 provides data summary statistics. Table A1 in the Appendix reports correlation analysis and test for multicollinearity results for relevant variables.

**Table 2.** Summary statistics

| Variables | Mean | Min | Max | Std. Dev. |
|---|---|---|---|---|
| Employment in all non-mining sectors (in 1000) | 1403.33 | 169.29 | 7546.96 | 1220.41 |
| Employment in the mining sector (in 1000) | 5.82 | 0.00 | 50.18 | 7.82 |
| Annual natural gas import from the US (million MCFs) | 606.29 | 53.13 | 1865.33 | 531.68 |
| Total state population (in 1000) | 3470.35 | 405.69 | 17753.90 | 2966.02 |
| Distance to the US-MX border (km) | 819.38 | 132 | 2032 | 499.97 |
| Price of imported natural gas (USD/MCF) | 4.33 | 2.04 | 8.25 | 1.70 |
| Study period | 1998-2019 | | | |
| Number of Mexican states | 32 | | | |
| Total number of observations | 704 | | | |

Note: (1) For states south of Mexico City, the distance to the US-MX border is computed as the distance to Mexico City plus the distance from Mexico City to the border. These states include Guerrero, Morelos, Puebla, Tlaxcala, Veracruz, Oaxaca, Chiapas, Tabasco, Campeche, Quintana Roo, Yucatán. (2) The natural gas price is the price for the pipeline-imported. Pipelines account for over 98% of the total Mexico natural gas import from the US in 2019.

Table 3 presents the results of the baseline model specifications. Columns OLS (ordinary least squares) and FE (fixed-effects) represent specifications without and with the state fixed effects, respectively. Mexican states have a lot of spatial heterogeneities in terms of economic development. Therefore, it is reasonable to prefer the FE specification here. We can make two general qualitative observations from the baseline results. First, the increase of natural gas import from the US has a significant positive employment impact in both the mining sector and non-mining sectors. Second, the closer to the US-Mexico border, the larger the employment impact of natural gas import (the quantity effect). To further analyze the results quantitatively, we move to the preferred IV regression results.

**Table 3.** The OLS and state fixed-effects (FE) estimation results of employment impacts (quantity effect)

|  | Non-mining Employment | | Mining Employment | |
| --- | --- | --- | --- | --- |
|  | OLS | FE | OLS | FE |
| Population (in 1000) | 408.29 | 446.09 | 0.44 | -0.09 |
|  | (0.00) | (0.00) | (0.00) | (0.88) |
| Natural gas import (million MCFs) | 96.00 | 76.92 | 2.45 | 2.71 |
|  | (0.00) | (0.00) | (0.00) | (0.00) |
| Distance to border (km) | 7.35 | - | 1.60 | - |
|  | (0.48) |  | (0.07) |  |
| Distance × Natural gas import | -0.04 | -0.04 | -0.01 | -0.01 |
|  | (0.03) | (0.00) | (0.04) | (0.00) |
| Fixed effects | None | State | None | State |
| $R^2$ | 0.99 | 0.99 | 0.04 | 0.73 |
| Number of observations | 704 | 704 | 704 | 704 |

Note: (1) Robust standard errors are used. *p-values* are reported in the parentheses. (2) For easy reporting, employment in the regression models is measured as the direct count (not in 1000). (3) The high $R^2$ statistics for the non-mining employment specifications are expected given the strong correlation between population and total employment.

This study chooses two instrumental variables derived from the exogenous natural gas import price: one-year-lagged natural gas price and predicted natural gas price (based on an autoregressive model with three lags, i.e., the AR (3) model). Including different lengths of lags reflects the various planning and decision-making horizons among natural gas importers (and consumers). Some importers may be more forward-looking than other importers. The relevance argument for the instruments is that natural gas import price is highly correlated with the import demand. As discussed earlier, one of the main reasons Mexico relies on the US natural gas supply is the much more expensive alternative of developing its reserve. The exogeneity argument for the instruments is that the lagged price and the predicted price can only affect domestic employment through changing the natural gas supply/demand. Otherwise, a change in the US natural gas export price is irrelevant. In a hypothetical extreme case where the US and Mexico do not trade at all, any price variations on the US side should not matter. Therefore, we can argue that the chosen instruments are exogenous to the model. One thing to note is that we exclude the contemporary natural gas import price from the instruments. This is to avoid any other potential empirical issues due to the simultaneity between price and quantity.

Table 4 presents the IV regression results for the OLS and FE specifications. Here we focus on the estimation results with FE specifications. The Wu-Hausman Exogeneity Tests suggest that we should reject the null hypothesis that natural gas import (the variable of concern) is exogenous at the 5% confidence level. The first-stage $F$ statistics suggest that the chosen instruments are highly relevant. The rule of thumb for detecting weak instruments is a first-stage $F$ statistic less than 10. In our case, the $F$ statistic (27.18) is significantly larger than 10. The test confirms the relevance of the chosen instruments.

**Table 4.** The instrumental variable (IV) regression estimation results of employment impacts

|  | **Non-mining Employment** |  | **Mining Employment** |  |
| --- | --- | --- | --- | --- |
|  | OLS | FE | OLS | FE |
| Population (in 1000) | 407.41 | 371.97 | 0.46 | 1.65 |
|  | (0.00) | (0.00) | (0.00) | (0.13) |
| Natural gas import (million MCFs) | 294.12 | 276.96 | -0.72 | -1.99 |
|  | (0.01) | (0.00) | (0.92) | (0.41) |
| Distance to border (km) | 113.67 | - | -0.10 | - |
|  | (0.05) |  | (0.98) |  |
| Distance × Natural gas import | -0.21 | -0.18 | 0.00 | 0.00 |
|  | (0.03) | (0.00) | (0.90) | (0.44) |
| *Fixed effects* | No | State | No | State |
| *Instrumented* | Natural gas import | | | |
| *Wu-Hausman Test (p-value)* | 0.06 | 0.00 | 0.65 | 0.05 |
| *Instrument variables* | One-year lagged natural gas price, predicted natural gas price | | | |
| *First-stage F statistic* | 12.48 | 27.18 | 12.48 | 27.18 |
| *$R^2$* | 0.99 | 0.99 | 0.02 | 0.71 |
| *Number of observations* | 704 | 704 | 704 | 704 |

Note: (1) Robust standard errors are used. *p-values* are reported in the parentheses. (2) For easy reporting, employment in the regression models is measured as the direct count (not in 1000). (3) The null hypothesis for the Wu-Hausman Test is that the instrumented variable (natural gas import) is exogenous.

Focusing on the columns of FE specification in Table 4, overall, the employment impact of natural gas import in the combined non-mining sector is statistically significant. The employment impact in the mining sector is insignificant. Specifically, and first, state population is strongly associated with non-mining employment level as expected. The key result here is that non-mining employment on average increases about 277 for a one-million-MCFs increase in annual natural gas import for a state located at the US-Mexico border (i.e., *Distance* = 0). The result is highly significant as the p-value suggests. So is the associated distance effect. To further see the differences in employment impact across different states, we need to incorporate the distance effect. Precisely, the estimated marginal impact of natural gas import is [276.96 − Distance to border × 0.18]. Table 5 summarizes the estimated employment impacts for selected states from north to south.

**Table 5.** Estimated marginal employment impact of natural gas import for selected states (quantity effect)

| State | Distance to border (km) | Estimated marginal effect | Standard error |
|---|---|---|---|
| Sonora | 184 | 243.27 | 41.77 |
| Tamaulipas | 209 | 238.69 | 40.81 |
| Durango | 488 | 187.61 | 30.23 |
| Aguascalientes | 610 | 165.27 | 25.77 |
| México | 788 | 132.68 | 19.64 |
| Oaxaca | 1188 | 59.44 | 11.55 |
| Chiapas | 1605 | -16.91 | 20.47 |
| Quintana Roo | 2032 | -95.10 | 35.95 |

Note: The standard errors are computed using the delta method. The unit of marginal effect is *jobs per million MCFs*.

The estimated marginal impacts in Table 5 suggest that the employment impact of natural gas import declines from north to south. Sonora is the highest, 243.27 jobs per million MCFs; Quintana Roo is the lowest, -95.10 jobs per million MCFs. The average employment impact across all 32 states is about 127 jobs per million MCFs of natural gas import. It is consistent with the network theory. As the pipeline network reaches the south, the network capacity requirement reduces. The associated regional economic impact hence also reduces. It is worth noting that a few southern states have negative employment impacts: Campeche, Chiapas, Quintana Roo, Tabasco, Yucatán (two of them are statistically significant: Quintana Roo and Yucatán). Although their estimated employment impacts are largely driven by the average-based regression model estimates ($\hat{\beta}_2$ and $\hat{\beta}_3$) in the equation (1), it is not simply a model artifact. In the past several decades, some of the poorest states have been in southern Mexico, including Guerrero, Chiapas, Oaxaca, and Veracruz. If we only consider the states south of Mexico City (11 states, see the footnote of Table 2 for the full list), the average estimated employment impact is about 30 jobs per million MCFs of natural gas import. It is substantially lower compared to the national average. We explore the policy implication of these findings in the discussion section.

*3.5 Price Effect - data and empirical results*

As previously discussed, the impact of natural gas imports on the combined non-mining sector works mainly through the price mechanism. To assess the price effect of natural gas trade on state-level employment (equation (2)), we need to collect additional data on state-level natural gas prices in Mexico. The natural gas price data comes from the "*Final price to the public of natural gas for the residential sector by geographical distribution area (without value-added taxes)*" reported by the Energy Information System (SIE) with information from the Energy Regulatory Commission (CRE, SENER) (SENER, 2021). It is referred to as commercial prices in SENER data documents. The data were collected from the 22 major cities in Mexico annually between 2000 and 2017. For states with one or more cities included in the raw data, we use the average as a state-level price estimate. For states with no cities included in the raw data, we estimate the state-level prices by spatially interpolating all bordering neighbor states with the simple average method. The prices are in nominal Pesos/Gigajoule (GJ) in the raw data. We deflated the prices into constant value using the Mexico GDP deflator (base year = 2015) provided by the OECD.[4] The raw price data has a mean of 143.16 Pesos/GJ and a standard deviation of 46.53 Pesos/GJ. The deflated price data has a mean of 182.39 Pesos/GJ and a standard deviation of 38.63 Pesos/GJ. Table A2 in the Appendix reports correlation analysis and test for multicollinearity results for relevant variables.

The natural gas price in Mexico is exogenous to its regional economy, as argued in sub-section 3.3. An additional piece of supporting evidence for the price exogeneity is the limited to none subsidy on domestic natural gas prices (Alcazar and Villalvazo, 2017; Martinez, 2021). Table 6 presents the estimation results for the new model (Equation (2)). Based on the preferred fixed effects (FE) model specifications, there are three main results: (1) natural gas (real) price changes have no impact on non-mining employment; (2) natural gas (real) price changes have a positive impact on mining employment, statistically significant at the 10% level; (3) there is a distance effect associated with the price effect. The

---

[4] The GDP deflator data is available through the US Federal Reserve Bank at St. Louis. Weblink for the Data: https://fred.stlouisfed.org/series/MEXGDPDEFQISMEI, accessed on Oct 2, 2021.

*p-value* for the distance interaction term is just at the 10% level. The negative estimate (-0.35) suggests that the price effect declines going from north to the south as expected. It is consistent with the quantity effect (Table 5). Quantitatively, the significant impact on mining employment implies that, for a one-percentage increase in natural gas price (1.82 Pesos/GJ, given the mean price at 182.39 Pesos/GJ), the state-level mining employment increases by 140.20 (approximately 2.38%, given the mean mining employment at 5,890.92 for the study period 2000-2017). The estimate 2.38 here also has an interpretation of the price elasticity of employment. The result suggests a sizeable regional economic impact on mining employment. Such an impact is consistent with the economic theory of substitution. As the imported natural gas becomes more expensive, the domestic mining sectors benefit from the growing demand for other energy sources (e.g., coal) substituting natural gas. We explore relevant policy implications of this result in the following discussion section.

**Table 6.** The OLS and state fixed-effects (FE) estimation results of employment impacts (price effect)

|  | **Non-mining Employment** | | **Mining Employment** | |
|---|---|---|---|---|
|  | OLS | FE | OLS | FE |
| *Log*(Population [in 1000]) | 0.99 (0.00) | 1.22 (0.00) | 0.48 (0.00) | 2.05 (0.00) |
| *Log*(Natural gas price [Peso/GJ]) | -0.16 (0.23) | -0.01 (0.91) | 2.26 (0.35) | 2.38 (0.07) |
| *Log*(Distance to border [km]) | -0.10 (0.36) | - | 0.72 (0.72) | - |
| *Log*(Distance) × *Log*(Natural gas price) | 0.02 (0.36) | -0.00 (0.73) | -0.18 (0.64) | -0.35 (0.10) |
| *Fixed effects* | None | State | None | State |
| $R^2$ | 0.98 | 0.99 | 0.14 | 0.76 |
| *Number of observations* | 576 | 576 | 566 | 566 |

Note: (1) Heteroskedasticity-robust standard errors are used. *p-values* are reported in the parentheses. (2) For easy reporting, employment in the regression models is measured as the direct count (not in 1000). (3) The study period is reduced to 2000-2017 due to data limitation related to natural gas prices. (4) Mining employment has missing values for a few state-year combinations.

## 4. Discussion

*4.1 Policy Implication*

Our empirical findings carry implications for Mexico's regional economic development from two aspects: energy policy and trade policy. Mexico's recent energy reform has opened the possibility of developing its shale gas reserve through a competitive market. However, the reform has not reached the expected outcome due to a lack of implementation and policy fluctuations from administration to administration. Meanwhile, Mexico's demand for natural gas has been growing and will continue to grow. To meet the demand, Mexico will likely continue relying on US natural gas export in the coming decade. The critical question is how Mexico should move forward in terms of energy development. Our empirical results suggest that the economic benefit of relying on cheap US natural gas is positive in the near-to-medium terms. Our empirical model assumes no significant structural change when identifying the regional economic impact of natural gas trade (both the quantity effect and the price effect). In the long term, however, we can no longer ignore the possibilities of technological advancement and structural changes. Therefore, the long-term economic consequence of the energy dependence on the US is uncertain. When projecting the long-term economic development outcome, it is necessary to consider two factors. First, it is reasonable to expect that the energy reform in Mexico will make progress. It will then fundamentally change the energy market landscape. Second, it is critical to factor in the transition to renewable energy. A gradual transition to renewable energy is an inevitable trend for both developed and developing economies. In that sense, overly relying on natural gas import will slow down the renewable energy transition in Mexico. However, the slowing-down effect is conditional on US energy policy.

This study concerns trade policy mainly regarding regional inequality. One of the implications of our results is that the regional economic benefit of natural gas import is spatially uneven. The northern states naturally get more pipeline construction projects and associated maintenance programs. They also likely host more distribution facilities and hubs. These economic activities create jobs. The southern states have fewer such economic development opportunities due to the fact that the network density and flow capacity reduce from north to south. A similar energy trade-related regional inequality issue has been raised in China in an inter-regional context (Sun et al., 2017). Of course, there are also historical reasons for the regional inequality of trade benefits across Mexican states. The closer to the US-Mexico border, the greater the potential economic benefits from US-Mexico trade. The state fixed effects in the model capture these historical locational effects. Overall, our results imply that it is often necessary to integrate trade policy and other regional development policies to reduce regional inequality. Methodologically, it is worth noting that our empirical model is a partial equilibrium analysis. It is impossible to tell the whole picture of the regional impact of energy trade from this study. Still, our proposed empirical method is easy to implement compared to other computational tools (e.g., computational general equilibrium models). It has fewer variables & parameters and hence fewer measurement errors. In addition, the estimated average employment impacts are easy to interpret, which may be desirable for certain policymaking purposes.

*4.2 Energy Security in Mexico*

The foremost energy-related challenge to Mexico's economy is whether it should establish energy independence. Geological studies have shown that Mexico has abundant natural gas reserves, especially the shale gas in the east (e.g., González, 2016). It means that Mexico has sufficient resources to pursue energy self-sufficiency, at least pertaining to natural gas. Meanwhile, as many US-based shale development studies show, a shale boom usually brings significant positive economic impacts in a region (e.g., Feyrer et al., 2017). Spatial and cross-industry spillover effects are often observed (e.g., Lee, 2015; Wang, 2020), which justifies shale development as a potential opportunity for long-term economic prosperity. The question is whether Mexico is missing an economic development opportunity for some of its historically stressed regions, especially regions in eastern and northeastern Mexico. Besides, studies have shown that US natural gas exports will likely maintain a competitive advantage for quite a long time (e.g., Bernstein et al., 2016). For instance, a recent US Geological Survey (USGS) study reveals that the production in the Permian Basin could last for another 20 - 30 years (USGS, 2018), which puts Mexico's policy consideration related to energy development and energy security in the long-term perspective. It also suggests that Mexico's energy dependency on US export may last for some time. It poses both a challenge and an opportunity for Mexico's energy transition.

An alternative way to think about energy security is from an energy supply insurance policy perspective (Winzer, 2012). Self-sufficiency in energy consumption is not necessarily an insurance policy against energy price fluctuations. It is particularly the case in the natural gas market considering the limited to none price subsidy in Mexico (Martinez, 2021). Between importing from the US and pursuing self-sufficiency, it is reasonable to expect that the former carries fewer supply chain and technology risks, at least in the near-to-medium term. Presently, the cost of developing shale gas in Mexico is higher than importing natural gas from the US for both market and technical reasons. In a competitive energy market like what is proposed in Mexico, pursuing energy independence seems to be an undesirable choice. It has led to a growing debate over energy dependence and security (e.g., Paraskova, 2019). A key question here is whether energy dependency is a vulnerability of regional economies in Mexico. This study provides some insights for answers. First, pipeline construction across the US-Mexico border should be a short-term economic development opportunity. The long-term dependency on US natural gas export will likely hurt Mexico's energy development and innovation capability for regional economic development. Second, the federal government, including national guards, police, and justice reform efforts, should improve the security of regions with natural gas deposits (e.g., Veracruz and Tamaulipas) and control local illegal activities. The automobile manufacturing industry in Mexico has taken similar measures, and they have proven effective. Lastly, the newly created regulatory entities (e.g., the CRE, SENER) should focus on efficient and transparent administrative processes to send a clear signal to the investors. For instance, these agencies should ensure effective enforcement of antitrust regulations to safeguard competitiveness and innovation in the market. The governments should also consider opening other sectors related to the natural gas industry to private investment, such as transportation and power generation.

## 5. Concluding Remarks

This study explores the regional economic impact of the US-Mexico natural gas trade in Mexico. We first reviewed the history and policy background related to the fossil energy sector in Mexico. We then developed a fixed-effects regression model to quantify the impact of the growing natural gas trade on Mexico's state-level employments. The model allows us to estimate the near-to-medium-term employment impact of the natural gas trade. We assessed both the quantity effect and the price effect of the dependency on natural gas import from the US. The quantity effect analysis suggests that natural gas import from the US has a significant positive impact on state-level non-mining employment. The estimated employment impact decreases from north to south, which can be explained by the diminishing network density and capacity going from north to south. Mexico's historical regional economic development inequality also contributes to the decline of the employment impact. We find no employment impact in the mining sector. It is because Mexico's mining sector has been small in recent decades. As far as the price effect is concerned, the empirical results suggest that natural gas price increases have a positive impact on mining employment but no statistically significant impact on non-mining employment. It can be potentially explained by the economic theory of substitution. As imported natural gas becomes more expensive, the domestic mining sectors benefit from the growing demand for other energy sources that substitute natural gas. There is also a similar distance effect associated with the price effect. It suggests that northern and central Mexican states benefit more from the trade-induced regional economic integration with the US than southern Mexican states.

We further explored the policy implications of our findings by focusing on energy policy and trade policy. Our findings are consistent with the classical trade theory. Mexico's regional economies mostly benefit from the growing natural gas trade. Nevertheless, considering the trade-off between short-term economic benefits and long-term economic prosperity, we suggest four strategies to move forward. The Mexican government should take actions to (1) grow innovation capacity to enable the development of its natural gas industry in the long term, although it may not be a cost-efficient strategy in the near-to-medium term; (2) attract investments and devote policy effort aiming at long-term energy security and economic development; (3) embrace the opportunities of renewable energy transition and sustainable energy development. Mexico was left behind during the shale revolution. There is no reason for Mexico to be left behind again in the coming renewable energy revolution. (4) Both Mexico's federal and local governments should effectively address the political and socio-economic uncertainties and create a healthy environment for business development and economic growth. It means that consistency in government policy across different administrations is critical. So is the continuing effort in policy implementation and regulation enforcement at different administrative levels. Our findings and discussion also shed light on the current energy security debate in Mexico.

Graham D (2020) Exclusive: U.S, Canada, European nations meet to discuss concern over Mexico energy policy. Available at https://www.reuters.com/article/us-mexico-energy-diplomacy-exclusive-idUSKBN20W0GI, accessed March 29, 2020.

Grayson G (1981) *The Politics of Mexican Oil*. University of Pittsburgh Press, Pittsburgh, PA.

Gross S (2019) AMLO reverses positive trends in Mexico's energy industry. *Order from Chaos*, Dec 20, 2019, the Brookings Institution. Available at https://www.brookings.edu/blog/order-from-chaos/2019/12/20/amlo-reverses-positive-trends-in-mexicos-energy-industry/, accessed on Oct 2, 2021.

Gutiérrez R (2014) Structural reforms in Mexico during the administration of Felipe Calderon: The energy one. *Economía UNAM*, 11(32): 32-58. Available at http://www.scielo.org.mx/pdf/eunam/v11n32/v11n32a3.pdf, accessed Dec 1, 2020.

Haahr K (2015) Addressing the concerns of the oil industry: Security challenges in Northeastern Mexico and government responses. Wilson Center Mexico Institute Working Paper, Washington, DC.

Ibarzábal JAH (2017) Examining governability of Mexico's natural gas transmission pipelines under the energy reform. *Journal of Energy & Natural Resources Law*, 35(3): 271-291.

Laguna-Martinez MG, Garibay-Rodriguez J, Rico-Ramirez V, Castrejon-Gonzalez EO, Ponce-Ortega JM (2020) Water impact of an optimal natural gas production and distribution system: An MILP model and the case-study of Mexico. *Chemical Engineering Research and Design*, 153: 887-906.

Lee J (2015) The regional economic impact of oil and gas extraction in Texas. *Energy Policy*, 87: 60-71.

Márquez M (1988) La industria del gas natural en México. *Problemas del Desarrollo*, 19(75): 39-67. DOI: www.jstor.org/stable/43907492.

Martínez RB (2021) Análisis de opinión del autotransporte de pasajeros en México, uso de energías alternativas y disminución de CO2. *Revista Internacional de Contaminación Ambiental*, 37: 389-400.

Navarro-Pineda FS, Handler R, Sacramento-Rivero JC (2017) Potential effects of the Mexican energy reform on life cycle impacts of electricity generation in Mexico and the Yucatan region. *Journal of Cleaner Production*, 164: 1016-1025.
27

**Appendix: Supplementary Figures and Tables**

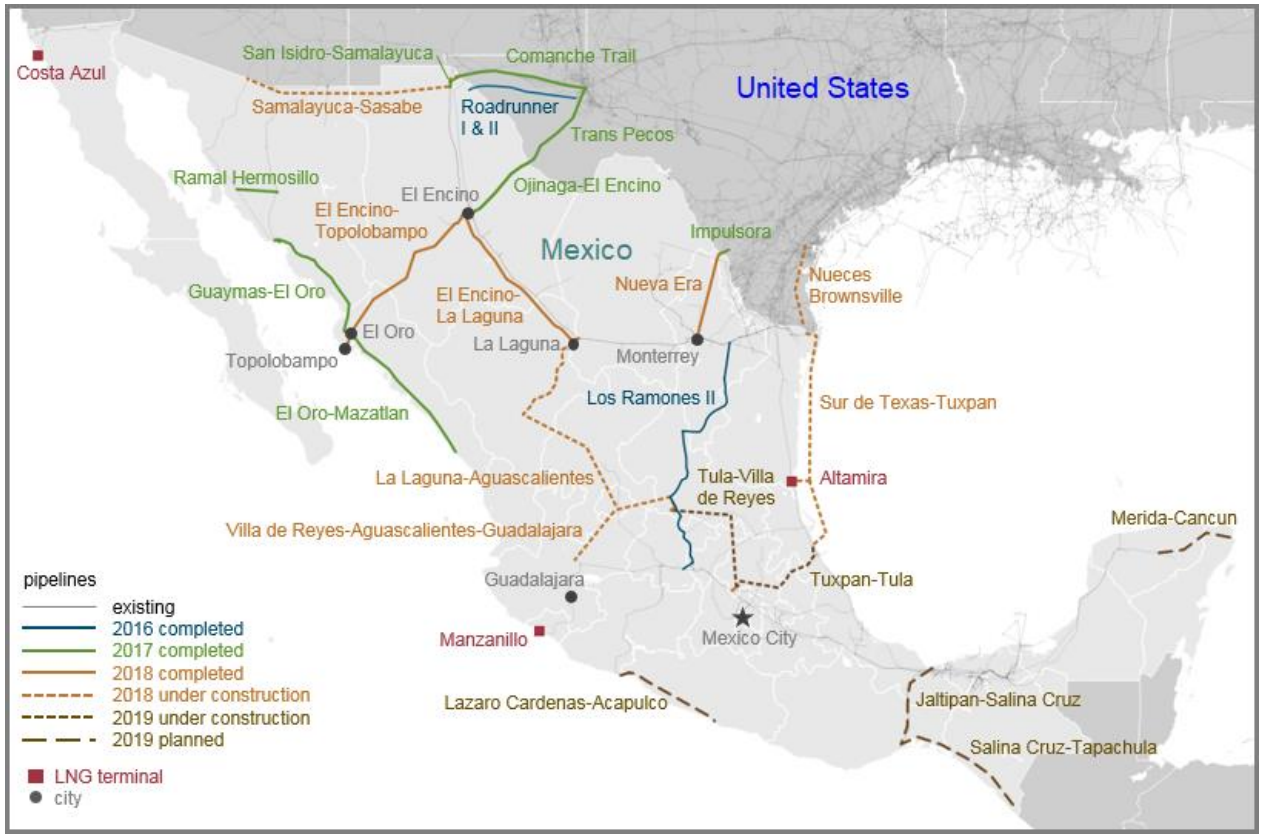

Data source: US EIA.

**Figure A1.** The recent expansions of Mexico's domestic natural gas pipeline network

**Table A1:** Correlation analysis and test for multicollinearity results for the quantity effect model

| | Dependent Variable: Non-mining Employment | | | |
|---|---|---|---|---|
| | Total employment | Total population | Natural gas import | Distance to border |
| Total employment | 1.0000 | | | |
| Total population | 0.9951 | 1.0000 | | |
| Natural gas import | 0.1125 | 0.0845 | 1.0000 | |
| Distance to border | -0.0680 | -0.0623 | 0.0000 | 1.0000 |
| VIF | | 1.0111 | 1.0072 | 1.0039 |
| | **Dependent Variable: Mining Employment** | | | |
| | Total employment | Total population | Natural gas import | Distance to border |
| Total employment | 1.0000 | | | |
| Total population | 0.1708 | 1.0000 | | |
| Natural gas import | 0.0685 | 0.0845 | 1.0000 | |
| Distance to border | 0.0138 | -0.0623 | 0.0000 | 1.0000 |
| VIF | | 1.0111 | 1.0072 | 1.0039 |

Note: Number of observations = 704.

**Table A2:** Correlation analysis and test for multicollinearity results for the price effect model

| | Dependent Variable: Non-mining Employment | | | |
|---|---|---|---|---|
| | Total employment | Total population | Natural gas price | Distance to border |
| Total employment | 1.0000 | | | |
| Total population | 0.9954 | 1.0000 | | |
| Natural gas price | -0.0565 | -0.0440 | 1.0000 | |
| Distance to border | -0.0672 | -0.0625 | -0.1039 | 1.0000 |
| VIF | | 1.0007 | 1.0135 | 1.0155 |
| | **Dependent Variable: Mining Employment** | | | |
| | Total employment | Total population | Natural gas price | Distance to border |
| Total employment | 1.0000 | | | |
| Total population | 0.1597 | 1.0000 | | |
| Natural gas price | 0.1548 | -0.0440 | 1.0000 | |
| Distance to border | 0.0446 | -0.0625 | -0.1039 | 1.0000 |
| VIF | | 1.0007 | 1.0135 | 1.0155 |

Note: Number of observations = 576.